\documentclass{epl}
\title{3D wedge filling and 2D random-bond wetting}
\author{J. M. Romero-Enrique\inst{1} \and A. O. Parry\inst{2}}
\institute{
\inst{1} Departamento de F\'{\i}sica At\'omica, Molecular y
Nuclear, Area de F\'{\i}sica Te\'orica, Universidad de Sevilla,
Apartado de Correos 1065, 41080 Sevilla, Spain\\
\inst{2} Department of Mathematics, Imperial College 180 Queen's Gate,
London SW7 2BZ, United Kingdom}
\pacs{68.08.Bc}{Wetting}
\pacs{05.70.Np}{Interface and surface thermodynamics}
\pacs{68.08.De}{Structure, measurements and simulations}
\begin{document}
\maketitle
\begin{abstract}
Fluids adsorbed in 3D wedges are shown to exhibit two types of continuous
interfacial unbinding corresponding to critical and tricritical filling
respectively. Analytic solution of an effective interfacial 
model based on the transfer-matrix formalism allows us to obtain the 
asymptotic probability distribution functions for the interfacial height 
when criticality and tricriticality are approached. Generalised random walk 
arguments show that, for systems with short-ranged forces, the critical 
singularities at these transitions are related to 2D complete and critical 
wetting with random bond disorder respectively. 
%The effect of long-ranged forces on the tricriticality is also 
%discussed. In particular, the tricritical point disappears for sufficiently 
%long-ranged forces.
\end{abstract}

Recent studies of filling transitions for fluids in 3D wedges \cite{Parry,
Parry2} have revealed the much stronger influence of interfacial fluctuations
compared with wetting at flat and rough substrates \cite{Dietrich, Bob, Schick,
Sartoni,Kardar}. Encouragingly
effective Hamiltonian predictions for the critical exponents at continuous 
(critical) wedge filling with short-ranged forces have been confirmed in 
large scale Ising model simulation studies \cite{Milchev}. 
Similar experimental verification of the predicted geometry-dominated 
adsorption isotherms at complete wedge filling \cite{Bruschi} raise hopes
that the filling transition itself and related fluctuation effects will be
observable in the laboratory. Here we further develop the fluctuation theory
of 3D filling and show that there is a rather deep and previously unrecognized 
connection with the theory of wetting in 2D systems with random bond (RB) 
disorder. Our findings are based on analytical solution of an effective model 
of 3D wedge filling and also generalised random-walk arguments \cite{Fisher}. 
First we show that there are actually two types of continuous filling behaviour 
corresponding to critical and tricritical transitions respectively with the 
latter having stronger fluctuation effects. The phase diagram for these 
transitions together with the classification of fluctuation regimes and the 
allowed values of critical exponents resemble very closely those predicted for
2D wetting. More precisely 3D critical filling is related to 2D complete
wetting whilst 3D tricritical filling is related to 2D critical wetting.
Remarkably the particular value of the 3D wedge wandering exponent for
pure systems (thermal disorder) implies that criticality at tricritical
and critical filling is  related to predictions for 2D critical
and complete wetting with RB disorder.

\begin{figure}
\onefigure[width=12cm]{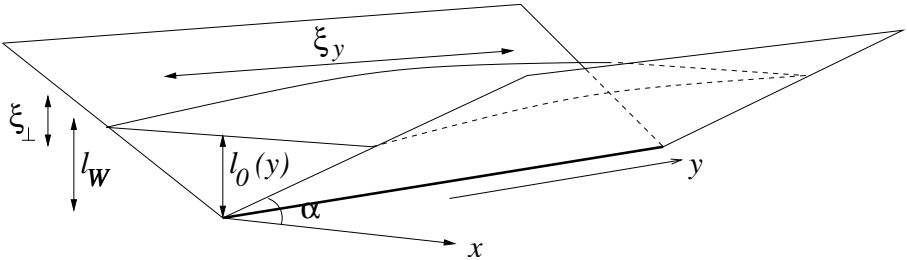}
\caption{Schematic illustration of a typical interfacial configuration
and relevant lengthscales for a fluid adsorption in a 3D wedge. The thick
line marks the position of the wedge bottom.}
\label{fig1}
\end{figure}

Consider the interface between a bulk vapour at temperature $T$ and
saturation pressure with a 3D wedge characterised by a tilt angle $\alpha$
(see Fig.~\ref{fig1}).
Macroscopic arguments dictate that the wedge is partially filled by liquid if 
the contact angle $\theta>\alpha$ and completely filled if $\theta< \alpha$ 
\cite{Concus} and is fully supported by both interfacial Hamiltonian \cite{RND}
and exact Ising studies \cite{Abraham}. The filling transition refers to the 
change from microscopic to 
macroscopic liquid adsorption as $T\to T_f$, at which $\theta(T_f)=\alpha$, and
may be first-order or continuous (critical filling). Both of these transitions 
can be viewed as the unbinding of the liquid-vapour interface from the wedge 
bottom. Characteristic length scales are the mean interfacial height above the 
wedge bottom $l_W$, the roughness $\xi_\perp$ and the longitudinal correlation 
length $\xi_y$, measuring fluctuations along the wedge (see Fig.~\ref{fig1}). 
The relevant scaling fields at critical filling are $\theta-\alpha$ and the 
bulk ordering field (partial pressure) $h$. In our discussion of filling we 
shall work exclusively at bulk coexistence ($h=0$) since it is here that the 
connection with RB wetting emerges. However calculations away from coexistence 
are not in any way problematic. At coexistence we define critical exponents
by $l_W\sim (\theta-\alpha)^{-\beta_W}$ and $\xi_y \sim (\theta-
\alpha)^{-\nu_y}$. The roughness can be related to $\xi_y$ by the
scaling relationship:
\begin{equation}
\xi_\perp\sim \xi_y^{\zeta_W}
\label{zw1}
\end{equation}
where $\zeta_W$ is the wedge wandering exponent.

The liquid-vapour interface across the wedge is aproximately flat and
soft-mode fluctuations arise from local translations in the height of
the filled region along the wedge \cite{Parry}. The pseudo-one-dimensional 
nature of these means that $\zeta_W$ is greater than the wandering exponent 
defined for the asymptotically flat free interface $\zeta_3$ in a 3D system. 
For systems with sufficiently short-ranged forces critical filling is
fluctuation-dominated (i.e. $\xi_\perp\sim l_W$) and dimensional
reduction arguments lead to the identification \cite{Parry2}:
\begin{equation}
\zeta_W=\frac{\zeta_2}{1+\zeta_2-\zeta_3}
\label{zw2}
\end{equation}
with $\zeta_2$ the 2D free interface wandering exponent. Thus for pure
(thermal) systems for which $\zeta_d=(3-d)/2$ for $d\le 3$, the 3D wedge
wandering exponent $\zeta_W=1/3$. In the fluctuation-dominated regime the
value of $\zeta_W$ determines the other exponents, in particular
\begin{equation}
\nu_y=\frac{1}{2-2\zeta_W}
\label{nuy}
\end{equation}
with $\beta_W=\nu_\perp=\zeta_W \nu_y$. Thus for thermal forces $\nu_y=3/4$, 
a value verified in Ising simulation studies \cite{Milchev}.

There is, however, another example of continuous filling with
even larger fluctuations characterized by different scaling fields and 
critical exponents. 
This corresponds to tricritical filling and there are two ways the transition 
may be induced. The first mechanism occurs for wedges made from walls which 
themselves exhibit weakly first-order wetting transitions \cite{Parry}. 
The filling transition in these systems will be first-order if $\alpha<
\alpha^*$ and critical for $\alpha>\alpha^*$, where $\alpha^*$ is the 
tricritical wedge angle. A second mechanism exists, however, which is more 
practicable in simulation studies. Imagine a wedge made from a homogeneous 
chemical material which exhibits critical filling. Now micropattern a stripe 
along the wedge bottom (see Fig.~\ref{fig1}), so as to weaken the local 
wall-fluid intermolecular potential and therefore strengthen the interfacial 
binding potential, since locally liquid adsorption will not be favoured and 
consequently the vapour-liquid interface is more likely to be localized near 
the wedge bottom. This situation can be easily engineered in Ising model 
studies by modifying the strength of the surface field near the wedge bottom 
and indeed has been done on planar substrates in the laboratory. With this 
modification it may be possible to bind the interface to the wedge bottom even 
at the filling boundary $\theta=\alpha$ and at bulk coexistence. A continuous 
tricritical transition may then be induced as the strength of the modified 
wedge potential approaches a tricritical value.

\begin{figure}
\onefigure[width=12cm]{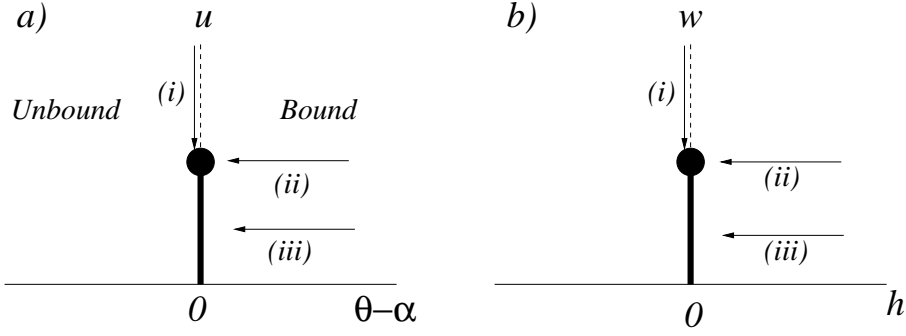}
\caption{Phase diagrams for (a) filling and (b) wetting transitions in terms
of the ordering fields $\theta-\alpha$ and $h$ and the contact-like potential 
strengths $u$ and $w$. The thick 
and dashed lines in both diagrams correspond to continuous and first-order
boundaries between bound and unbound interfacial states, respectively.
The arrows show representative paths along which continuous unbinding
occur. The filled circles represent the tricritical filling and critical 
wetting points, respectively.}
\label{fig2}
\end{figure}

For thermal systems, both critical and tricritical transitions can be
modelled by the wedge Hamiltonian \cite{Parry}
\begin{equation}
{\mathcal H}_W[l_0]=\int d y\left\{\frac{\Sigma l_0}{\alpha}
\left(\frac{dl_0}{dy}\right)^2
+V_W(l_0)\right\}
\label{heff2}
\end{equation}
where $l_0(y)$ is the local height of the interface at position $y$ along
the wedge bottom and $\Sigma$ is the liquid-vapor surface tension. This 
expression arises from the identification of the interfacial breather modes 
which translate the interface up and down the sides of the confining geometry,
as the only relevant fluctuations for the critical filling transition, and
corresponds to the excess free-energy contribution of a breather-mode
configuration with respect to the planar case obtained from the usual
capillary wave model \cite{Parry}. Note that the effective bending term 
resisting fluctuations along the wedge is proportional to the local interfacial
height since the interfacial cross-section has a length $2l_0/\alpha$. At bulk 
coexistence, the effective binding potential $V_W(l_0)$, up to unimportant 
additive constants, is given by \cite{Parry}:
\begin{equation}
V_W(l_0)\approx \frac{\Sigma(\theta^2-\alpha^2) l_0}{\alpha}+
\Delta V_W(l_0)
\label{effbinding}
\end{equation}
where $\Delta V_W (l_0)$ has a hard wall repulsion for $l_0<0$ and a
long-ranged tail which decays when $l_0\to \infty$. For short-ranged forces
this can be modelled as a contact-like potential with strength $u$ (i.e. 
$\Delta V_W \sim -u \delta(l_0)$, where $\delta(x)$ is Dirac's delta function).
In general, there will be a tricritical value $u_c$ such that for $u<u_c$ the
interface unbinds from the wedge bottom when $\theta\to \alpha$ whilst for 
$u>u_c$, the interface remains bound to the wall in the same limit.
A section of the phase diagram (at $h=0$) for this is presented in 
Fig.~\ref{fig2}(a) and shows two continuous filling transitions. Critical 
filling corresponds to $\theta\to \alpha$ for $u<u_c$ (route (iii)) and has 
the critical exponents described above. Tricritical filling corresponds to any 
thermodynamic path for which $u\to u_c$ and $\theta\to \alpha$ (routes (i) and 
(ii), for example). Along the path $\theta=\alpha$ the relevant length scales 
diverge as:
\begin{equation}
l_W\sim (u-u_c)^{-\beta_W^*} \ ,\ \xi_y\sim (u-u_c)^{-\nu_y^*}
\label{critexp1}
\end{equation}
and $\xi_\perp\sim \xi_y^{\zeta_W^*}$, with $\zeta_W^*$ the tricritical 
wandering exponent. These expressions define new critical
exponents which are distinct from those at critical filling.
Again for short-ranged forces we anticipate that the transition is
fluctuation-dominated with $l_W\sim \xi_\perp$ and $\beta_W^* = \zeta_W^*
\nu_y^*$. More generally, in the vicinity of the tricritical point (and
at $h=0$) we anticipate scaling e.g. $\xi_y\sim |u-u_c|^{-\nu_y^*} \Lambda
\left[(\theta-\alpha)|u-u_c|^{-\Delta^*}\right]$ with the gap exponent
$\Delta^*$. Thus along route (ii) $\xi_y\sim (\theta-\alpha)^{-\nu_y^*/
\Delta^*}$.

To continue we evaluate these exponents for thermal systems by explicit
transfer-matrix analysis before recasting them more generally in terms of
the wandering exponent $\zeta_W$. In the continuum limit the partition function
is defined as a path integral \cite{Burkhardt}:
\begin{equation}
Z[l_b,l_a,Y]=\int {\mathcal D}l_0 \exp(-{\mathcal H}_W[l_0])
\label{partfunc}
\end{equation}
where $Y$ is the wedge length, $l_a=l_0(0)$ and $l_b=l_0(Y)$ are the endpoint 
heights
and we have set $k_B T=1$ for convenience. Due to the presence of a
position-dependent stiffness some care must be taken with the
definition of the path integral. This turns out to be of crucial
importance for the the evaluation of the exponents at tricritical (but
not critical) filling. This problem was already pointed out in 
Ref.\cite{Bednorz} and is related to the well-known ordering
problem in the quantization of classical Hamiltonians with
position-dependent masses. Similar issues also arise in solid state physics 
\cite{Thomsen}. Borrowing from the methods used to overcome these difficulties
we use the following definition
\begin{equation}
Z[l_b,l_a,Y]=\lim_{N\to \infty} \int dl_1\ldots dl_{N-1} \prod_{j=1}^N
K(l_j,l_{j-1},Y/N)\label{partfunc2}
\end{equation}
where $l_0\equiv l_a$ and $l_N\equiv l_b$, and $K(l,l',y)$ is defined as:
\begin{equation}
K(l,l',y)=\sqrt{\frac{\Sigma\sqrt{ll'}}{\alpha \pi y}}
\exp\left(-\frac{\Sigma \sqrt{ll'}}{\alpha y}(l-l')^2-
y V_W(l)\right)
\label{partfunc3}
\end{equation}
In the continuum limit the partition function becomes
\begin{equation}
Z(l_b,l_a,Y)=\sum_n \psi_n(l_b)\psi_n^*(l_a)
\textrm{e}^{-E_n Y}
\label{partfunc4}
\end{equation}
where the complete orthonormal set of functions satisfies
\begin{equation}
\Bigg(-\frac{\alpha}{4\Sigma}\frac{\partial}
{\partial l} \left[\frac{1}{l}\frac{\partial}{\partial l}\right]
+
V_W(l)-\frac{3\alpha}{16\Sigma l^3}\Bigg)\psi=E\psi
\label{schrodinger1}
\end{equation}
In the thermodynamic limit $Y\to \infty$ we obtain the probability
distribution function (PDF) for the midpoint interfacial height $P_W(l_0)=
|\psi_0(l_0)|^2$ and the longitudinal correlation length $\xi_y=
1/(E_1-E_0)$. We have obtained the analytical solution to the transfer-matrix 
operator for short-ranged forces determining the crossover from tricritical 
to critical filling. We present without proof some of our findings. A more
complete and detailed description of our results will be published elsewhere.

\begin{figure}
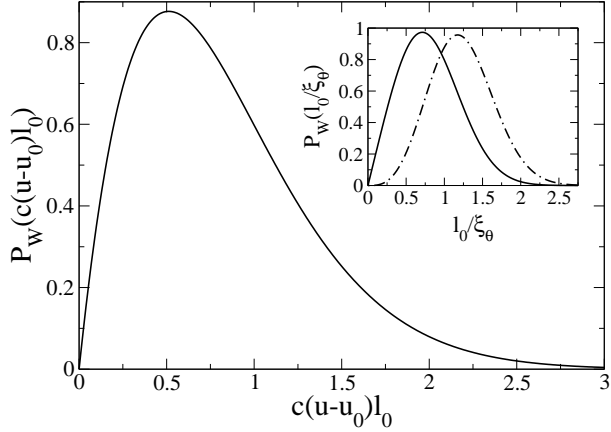

\onefigure[width=8cm]{fig3.eps}
\caption{Scaled interfacial height PDF $P_W(c(u-u_c)l_0)$ as a
function of the scaled wedge midpoint interfacial height $c(u-u_c)l_0$ for
$\theta=\alpha$. Inset: Plot of the scaled PDF along
routes (ii) and (iii) in Fig.~\ref{fig2}(a), i.e.
for $\epsilon\approx 1.086$ (continuous line) and $\epsilon\approx
1.639$ (dot-dashed line), respectively.} 
\label{fig3}
\end{figure}

Along route (i) we find that there is only one bound 
solution to Eq.~(\ref{schrodinger1}) for $u>u_c$ with $E_0\propto (u-u_c)^3$ 
and associated PDF
\begin{equation}
P_W(l_0)\propto l_0 \textrm{Ai}^2[c(u-u_c) l_0]
\label{PDF}
\end{equation}
with $c$ an unimportant constant while $\textrm{Ai}(x)$ is the Airy function. 
Thus $l_W\sim \xi_\perp \propto (u-u_c)^{-1}$ and $\xi_y\propto
(u-u_c)^{-3}$ identifying $\beta_W^*=1$, $\nu_y^*=3$ and confirming that
the tricritical wandering exponent coincides with the critical wandering
exponent $\zeta_W^*=\zeta_W=1/3$. A numerical plot of the PDF is shown in 
Fig.~\ref{fig3}. On the other hand, the scaling of the PDF along the route 
(ii) is given by
\begin{eqnarray}
P_W(l_0) \propto l_0 \exp\left[
\frac{2 \epsilon l_0}{\xi_\theta} - \frac{2l_0^2}{\xi_\theta^2}
\right]
H_s^2\left(\sqrt{2}\frac{l_0}{\xi_\theta}
-\frac{\epsilon}{\sqrt{2}}\right)
\label{PDF2}
\end{eqnarray}
where $\xi_\theta=\Sigma^{-1/2}[(\theta/\alpha)^2-1]^{-1/4}$,
$s=\epsilon^2/4-1/2$ with $\epsilon=-\Sigma E_0 \xi_\theta^3/\alpha
\approx 1.086$ and $H_s(z)$ is the Hermite function \cite{Lebedev}. The value 
of $\epsilon$ is obtained by imposing appropriate boundary conditions at 
$l_0=0$. Thus along this route $l_W\sim \xi_\perp \propto 
(\theta-\alpha)^{-1/4}$ similar to 
critical filling. From analysis of the spectrum it is also possible to show 
that $\xi_y\propto (\theta-\alpha)^{-3/4}$. As anticipated, in the vicinity of 
the tricritical point the divergent length scales show scaling with tricritical 
gap exponent $\Delta^*=4$. Whilst the exponents for critical filling are
already known the exact scaling form for the PDF has not been given previously.
For thermodynamic paths (iii) far from the tricritical point we have found that
the scaling of the PDF is of the form shown in Eq.~(\ref{PDF2}) but with 
$\epsilon\approx 1.639$. We remark that the PDFs at critical and tricritical 
filling have distinct short-distance expansions when $l_0/l_W\to 0$ and our 
results (cubic and linear powers respectively) are consistent with exact 
thermodynamic requirements \cite{Parry2}.

We can now place the results for tricritical filling in
a more general context. For short-ranged forces the tricritical transition
belongs to a general class of strong-fluctuation regime interfacial unbinding 
since the transition occurs at a finite value of the binding potential $u$.
The critical singularities at such transitions can be very elegantly modelled
using random-walk methods which previously have been succesfully used to
understand 2D critical wetting (at flat walls) \cite{Fisher}. By modelling the
interface as a sequence of bound and unbound regions (the so-called
bead-necklace picture), the critical exponent for the correlation length
along the interface at the strong-fluctuation transition can be related to
the appropriate wandering exponent. Details will be published elsewhere, so
we just mention our main results. Following this argument for the
tricritical filling transition and making allowance for the position-dependent
stiffness, we find
\begin{equation}
\nu_y^*=\frac{1}{1-2\zeta_W}\ ,\ \Delta^*=\frac{2-2\zeta_W}{1-2\zeta_W}
\label{bead}
\end{equation}
from which all other critical exponents follow. For thermal systems,
$\zeta_W=1/3$, implying $\nu_y^*=3$ and $\Delta^*=4$, in agreement with
our explicit calculations. Eqs.~(\ref{PDF}), (\ref{PDF2}) and (\ref{bead})
are the main results of our paper and together with Eq.~(\ref{nuy}) completely
determine the critical singularities at fluctuation-dominated 3D filling
occuring at bulk coexistence.

A remarkable connection with the theory of 2D complete and critical wetting
is now apparent. These transitions correspond to the continuous unbinding
of an interface from a planar wall and can be modelled by the interfacial
Hamiltonian
\begin{equation}
{\mathcal H}[l]=\int dx \left[\frac{\tilde \Sigma}{2}\left(\frac{dl}{dx}
\right)^2+W(l)\right]
\label{heff1}
\end{equation}
where $l(x)$ is the interfacial height at a position $x$ along the wall,
$\tilde \Sigma$ is the 2D stiffness and $W(l)$ is the binding potential.
In general $W(l)=hl+\Delta W(l)$ where $h$ is proportional to the bulk
ordering field (partial pressure) and for short-ranged forces $\Delta W(l)$
can be modelled as a contact potential of strength $w$ 
(i.e. $\Delta W(l)=-w\delta (l)$). Disorder arising
from bulk random bonds can also be allowed for by including a stochastic
term in $W(l)$. The phase diagram is shown in Fig.~\ref{fig2}(b), and
shows two continuous transitions referred to as complete wetting (path 
(iii)) and critical wetting (e.g. path (i) and (ii)), at which
the mean interfacial height $l_\pi$, roughness $\xi_\perp$ and transverse
correlation length $\xi_\parallel$ all diverge. For short-ranged forces the 
transitions are fluctuation-dominated and $l_\pi \sim \xi_\perp\sim 
\xi_\parallel^{\zeta_2}$. More specifically, for complete wetting we write 
$\xi_\parallel \sim h^{-\nu_\parallel^{co}}$
whilst for critical wetting $\xi_\parallel \sim (w-w_c)^{-\nu_\parallel}$
with associated gap exponent $\Delta$ off coexistence.
The values of these exponents can be expressed explicitly in terms of
the wandering exponent $\zeta_2$ as \cite{Lipowsky}
\begin{equation}
\nu_\parallel^{co}=\frac{1}{2-\zeta_2}\ ,\
\nu_\parallel=\frac{1}{1-\zeta_2}\ ,\
\Delta=\frac{2-\zeta_2}{1-\zeta_2}
\label{wetting}
\end{equation}
There is therefore a qualitative and quantitative connection between critical
singularities at 3D filling and 2D wetting. The phase diagrams are equivalent 
(see Fig.~\ref{fig2}) with the field $\theta-\alpha$ playing the role of the 
ordering field $h$ at wetting. Further, writing the 2D wetting exponents in 
terms of $\zeta_2$, e.g. $\nu_\parallel^{co}=\nu_\parallel^{co}(\zeta_2)$, 
etc., we have from Eqs.~(\ref{nuy}) and (\ref{bead}) the dimensional reduction 
relations for the critical exponents
\begin{equation}
\nu_y=\nu_\parallel^{co}(2\zeta_W)\ ,\ \nu_y^*=\nu_\parallel(2\zeta_W)\ ,
\ \Delta^*=\Delta(2\zeta_W)
\label{mapping}
\end{equation}
Most remarkably for thermal forces the numerical value $2\zeta_W=2/3$ means 
that 3D filling is related to 2D wetting with RB disorder, since $\zeta_2=2/3$ 
for this case \cite{Huse}.

%Finally we mention that the interfacial model Eq.~(\ref{heff2}) can be
%also used to study wedge filling with long-ranged forces.
%Allowing for long-ranged attractive $-a/l^p$ and repulsive $b/l^q$ ($q>p$) 
%algebraic tails in the 3D interface-planar substrate binding potential
%$W(l)$, two different scenarios arise. If $p>4$, the long-ranged interactions 
%only renormalize the short-distance substrate-interface interaction and 
%the tricritical behavior, as well as the critical filling, remains unchanged.
%This situation corresponds to the fluctuation-dominated regime. On the other 
%hand, $\Delta V_W(l_0)$ develops a repulsive tail as $l_0 \to 
%\infty$ if $p<4$ and the tricritical point disappears. For van der Waals forces 
%($p=2$), explicit calculations confirm this scenario: along the filling 
%transition boundary, a first-order interfacial unbinding transition 
%with a diverging $\xi_y$ is observed instead of the tricritical point. 

A full account of our study, including the analysis of the presence of 
long-ranged forces, will be presented elsewhere.

In summary we have identified a second example of continuous filling transition
corresponding to tricritical behaviour. For systems with short-ranged forces
and thermal disorder only we have exactly found the critical singularities
and associated probability distribution function for the interfacial height. 
A random-walk analysis
reveals a remarkable connection between the critical exponents for thermal 
3D filling and random-bond 2D wetting for systems with short-ranged forces. 
These predictions may certainly be tested in Ising model simulation studies 
and would be a stringent test of the theory of 3D wedge filling.
Experimental studies of tricritical (and critical) filling similar to those 
already performed for complete filling would be very welcome.

\acknowledgements
J.M.R.-E. acknowledges financial support from the European Commission
under Contract MEIF-CT-2003-501042.


\begin{thebibliography}{199}
\bibitem{Parry} 
\Name{Parry A. O., Rasc\'on C. \and Wood A. J.}
\REVIEW{Phys. Rev. Lett.}{85}{2000}{345};
\Name{Parry A. O., Wood A. J.\and Rasc\'on C.}
\REVIEW{J. Phys.: Condens. Matter}{13}{2001}{4591}
\bibitem{Parry2} 
\Name{Greenall M. J., Parry A. O. \and Romero-Enrique J. M.}
\REVIEW{J. Phys.: Condens. Matter}{16}{2004}{2515}
\bibitem{Dietrich}
\Name{Dietrich S.} in
\Book{Phase Transitions and Critical Phenomena}
\Vol{12}
\Editor{C. Domb \and J. L. Lebowitz}
\Year{1988} 
\Publ{Academic Press, London}
\bibitem{Bob}
\Name{Evans R.} in
\Book{Liquids at Interfaces (Les Houches Sessions XLVIII)}
\Editor{J. Charvolin, J. F. Joanny and J. Zinn-Justin}
\Year{1990}
\bibitem{Schick}
\Name{Schick M.} in
\Book{Liquids at Interfaces (Les Houches Sessions XLVIII)}
\Editor{J. Charvolin, J. F. Joanny and J. Zinn-Justin}
\Year{1990}
\bibitem{Sartoni}
\Name{Sartoni G., Stella A. L., Giugliarelli G. \and D'Orsogna M. R.}
\REVIEW{Europhys. Lett.}{39}{1997}{633}
\bibitem{Kardar}
\Name{Kardar M. \and Indekeu J. O.}
\REVIEW{Europhys. Lett.}{12}{1990}{161}
\bibitem{Milchev} 
\Name{Milchev A., M\"uller M., Binder K. \and Landau D. P.}
\REVIEW{Phys. Rev. Lett.}{90}{2003}{136101}
\bibitem{Bruschi} 
\Name{Bruschi L., Carlin A. \and Mistura G.}
\REVIEW{Phys. Rev. Lett.}{89}{2002}{166101}
\bibitem{Fisher} 
\Name{Fisher M. E.}
\REVIEW{J. Chem. Soc. Faraday Trans. 2}{82}{1986}{1589};
\Name{Fisher M. E.}
\REVIEW{J. Stat. Phys.}{34}{1984}{667}
\bibitem{Concus} 
\Name{Concus P. \and Finn R.}
\REVIEW{Proc. Natl. Acad. Sci. USA}{63}{1969}{292}
\bibitem{RND} 
\Name{Rejmer K., Dietrich S. \and Napi\'orkowski M.}
\REVIEW{Phys. Rev. E}{60}{1999}{4027}
\bibitem{Abraham}
\Name{Abraham D. B. \and Maciolek A.}
\REVIEW{Phys. Rev. Lett.}{89}{2002}{286101}
\bibitem{Burkhardt} 
\Name{Burkhardt T. W.}
\REVIEW{Phys. Rev. B}{40}{1989}{6987}
\bibitem{Bednorz}
\Name{Bednorz A. \and Napi\'orkowski M.}
\REVIEW{J. Phys. A: Math. Gen.}{33}{2000}{L353}
\bibitem{Thomsen} 
\Name{Thomsen J., Einevoll G. T. \and Hemmer P. C.}
\REVIEW{Phys. Rev. B}{39}{1989}{12783}; 
\Name{Chetouani L., Dekar L. \and Hammann T. F.}
\REVIEW{Phys. Rev. A}{52}{1995}{82}
\bibitem{Lebedev} 
\Name{Lebedev N. N.}
\Book{Special Functions and their applications} 
\Publ{Dover Publications Inc., New York}
\Year{1972}
\bibitem{Lipowsky}
\Name{Lipowsky R. \and Fisher M. E.}
\REVIEW{Phys. Rev. Lett.}{56}{1986}{472}
\bibitem{Huse} 
\Name{Huse D. A., Henley C. L. \and Fisher D. S.}
\REVIEW{Phys. Rev. Lett.}{55}{1985}{2924}
\end{thebibliography}
\end{document}